# Relaxation and freezing of dielectric response in relaxor $Pb(Mg_{1/3}Nb_{2/3})O_3$ crystal

A. A. Bokov and Z.-G. Ye
*Department of Chemistry, Simon Fraser University, Burnaby, BC, V5A 1S6, Canada*

Dielectric investigations of the relaxor ferroelectric $Pb(Mg_{1/3}Nb_{2/3})O_3$ crystal were performed in the frequency domain at $10^{-4}$–$10^5$ Hz. In the ergodic relaxor phase two overlapping relaxation processes obeying the Kohlrausch-Williams-Watts law, $\exp[-(t/\tau)^\beta]$, and the Curie–von Schweidler law, $t^{-n}$, respectively, were found and parameterized. The values of $\beta$, $n$ and $1/\tau$, tend to zero at 211 K following the Vogel-Fulcher law, pointing to the glassy freezing of the dipole dynamics related to both relaxation processes.

Relaxation processes in many electric, magnetic, mechanical and other systems are governed by the Kohlrausch-Williams-Watts (KWW) law, $P \propto \exp[-(t/\tau)^\beta]$ (where $\tau$ is the relaxation time and $0<\beta\leq 1$ is the parameter). It describes the approach of a physical quantity $P$ (polarization in case of dielectric relaxation) to the equilibrium with time ($t$) [1]. Another general type of dielectric relaxation can be described by the Curie–von Schweidler (CS) law, $dP/dt \propto t^{-n}$ (where $0<n<1$ is the parameter) [1]. Although the microscopic nature of these laws remains puzzling, they are empirically well established.

The relaxation in highly disordered systems such as glasses is of particular interest. In many cases it does not follow the classical Arrhenius behavior. Instead, on cooling $\tau$ tends to infinity at a non-zero temperature $T_\tau$ which can be considered as the freezing temperature below which the system becomes nonergodic. To describe the temperature dependence of $\tau$ at $T > T_\tau$ the Vogel-Fulcher (VF) law is often used [2]:

$$\tau = \tau_0 \exp[E_\tau/(T - T_\tau)], \qquad (1)$$

where $\tau_0$ and $E_\tau$ are the parameters. This law was confirmed in many glass-forming materials (simple liquids and polymers) [2], some orientational (dipole and quadrupole) glasses [3], spin glasses [4] and relaxor ferroelectrics [5,6] (which can be considered as spherical cluster glasses [7]).

The relaxation in glasses and the origin of the VF law have been discussed in many works. For example, Monte Carlo simulation based on the Potts spin glass model revealed [8] an almost linear decrease of $\beta$ on cooling so that it becomes very small as $T\rightarrow 0$. In Ref. [2] the VF law (1) was deduced on the basis of the assumption that $\beta$ varies as $-1/T$. Thus for experimental verification of the theoretical models knowledge of the temperature dependences of relaxation parameters is desirable.

Experimentally the behavior of $\beta$ was studied elaborately in glass formers. It was found to decrease linearly with decreasing $T$ and the extrapolated temperature at which it becomes zero does not usually differ much from $T_\tau$ [2]. A decrease of $\beta$ on cooling was found in $KTaO_3$:Li dipole glass [9].

In the present work we analyze the low-frequency dielectric spectra in $Pb(Mg_{1/3}Nb_{2/3})O_3$ (PMN) crystal which is known as a prototypical relaxor ferroelectric. We derive the set of equations capable of fully describing the linear dielectric response in the ergodic relaxor phase (i.e. at $T > T_\tau$) and determine the parameters of these equations for PMN crystal. Overlapping CS and KWW relaxation processes are found. The KWW parameter $\beta$ decreases upon cooling according to the VF law and tends to zero at the same temperature where the extrapolated $\tau$ diverges. To the best of our knowledge, such kind of $\beta(T)$ behavior has never been reported so far for any material. Furthermore, the parameter $n$ of the CS relaxation also tends to zero according to the VF law at approximately the same temperature.

The relative dielectric permittivity $\varepsilon^* = \varepsilon' - i\varepsilon''$ was measured using impedance analyzer (Novocontrol turnkey dielectric spectrometer Concept 20). Gold electrodes were sputtered on the large faces of (100) oriented crystal plate of 0.16 mm thick grown as reported in Ref. 10. During cooling at 1 K/min from 640 K the temperature was stabilized (to within ±0.01 K) every 3 K (at $T$>325 K larger steps were used) and isothermal measurements of $\varepsilon^*$ were performed with a measurement ac field of 1 V/mm in the frequency range of $f=10^{-2}$-$10^{5}$ Hz (a wider range at selected temperatures). The average cooling rate at $T$<325 K was 0.1 K/min.

The measured $\varepsilon'(T)$ dependences (see Fig. 1) are similar to those reported for PMN crystal in previous works [5,11]. Strong dispersion at the low-$T$ side of the $\varepsilon'(T)$ peak leads to the displacement of the maximum temperature $T_{mr}$ with $f$. The dispersion at the high-$T$ side is much less pronounced but still exists, as can be seen in Fig. 2 that shows the $\varepsilon'(f)$ and $\varepsilon''(f)$ dependences. Two loss processes can be noticed. The first one gives rise to the $\varepsilon''(f)$ maximum which is clearly seen at $T$~250 K and moves out of the measurement frequency window upon increasing or decreasing $T$. The susceptibility related to the second (low-$f$) process grows monotonically with decreasing $f$ in the whole frequency range studied. Thus the total permittivity can be expressed as

$$\varepsilon^*(f,T) = \chi_U^*(f,T) + \chi_R^*(f,T) + \varepsilon_\infty(T)\,, \qquad (2)$$

where $\chi^*_U$ and $\chi^*_R$ are the susceptibilities related to the above mentioned low- and high-$f$ dispersion branches, respectively, and $\varepsilon_\infty$ represents the contribution of other possible polarization processes which have the dispersion at frequencies higher than the upper limit of measurement frequencies. Two similar dispersion regions were found before in the (1-x)$Pb(Mg_{1/3}Nb_{2/3})O_3$–x$PbTiO_3$ relaxor solid solution with x~0.3 (PMNT) [12,13] and the related parts of permittivity were called the "universal relaxor" ($\chi^*_U$) and the "conventional relaxor" ($\chi^*_R+\varepsilon_\infty$) contributions.

The universal relaxor (UR) dispersion can be readily analyzed at comparatively high $T$ where it dominates. Similar to PMNT this dispersion is found to follow the fractional power law,

$$\chi'_U(f) = \chi_{U1} f^{n-1} = \tan(n\pi/2)\,\chi''_U(f), \qquad (3)$$

where $n$ and $\chi_{U1}$ are the parameters. As is known, this formula represents the CS relaxation in the frequency domain [1]. The $\chi''_U(f)$ dependences in log-log presentation and the $\chi''(\chi'_U)$ dependences (Cole-Cole plots) obeying Eq. (3) must be the straight lines the slopes of which are determined by $n$. Such kind of linearity is evident from the high-$T$ plots in Fig. 2 (b) and from the Cole-Cole plots, one of which is shown in Fig. 3. As expected, the $n$ values obtained from both representations are found to be the same.

To fit the experimental data at given $T$ in the whole measurement frequency range with Eq. (2), we apply the nonlinear weighted least-square method. The relations (3) are used to model the first term of Eq. (2). To represent the second term we use one of the two forms: the frequency-domain transform of the KWW law or the well-known Havriliak-Negami (HN) relationship [1], $\chi_R^*(f) = \chi_{R0}[1+(i2\pi f\tau)^{\alpha}]^{-\gamma}$, where $0<\alpha<1$ and $\gamma>0$ are the parameters and $\chi_{R0}$ is the static susceptibility. Since the KWW relaxation in frequency domain is expressed by the Fourier transform of the derivative of the KWW function [1], one has:

$$\chi_R^*(f) = \chi_{R0}\beta\tau^{-\beta}\int_0^\infty t^{\beta-1}\exp[-(t/\tau)^{\beta} - i2\pi f t]\,dt\,. \qquad (4)$$

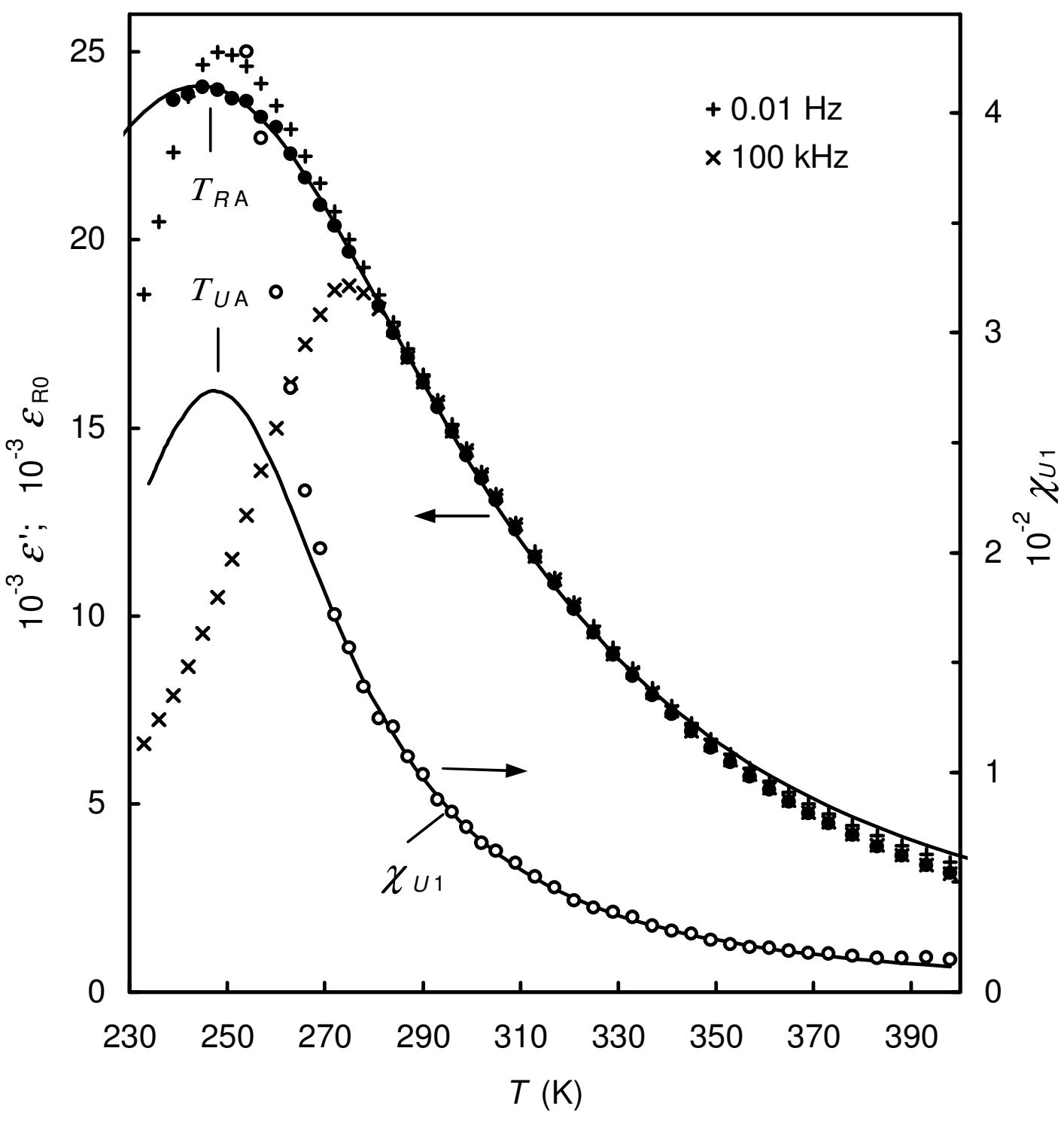

FIG. 1. Temperature dependences of the measured $\varepsilon'$ at $10^{-2}$ and $10^{5}$ Hz (crosses), the calculated static CR permittivity, $\varepsilon_{R0}=\chi_{R0}+\varepsilon_{\infty}$ (filled circles) and the calculated UR susceptibility, $\chi_{U1}$ (open circles) in PMN crystal. Solid lines are the fits to Eqs. (9), (10).

However, it is known that the analytic expression for this integral does not exist, which makes the routine least-squares algorithm impossible. Thus in the first step of the fitting procedure we use the approximate analytic formulae of $\varepsilon^{*}(f)$ for the KWW relaxation [14]. In the second step, when approximate values of adjustable parameters are known, we fit experimental data directly with numerically calculated integral (4) using the trial-and-error procedure and the least-squares as an estimate for goodness of the fit. To enhance the reliability, in all steps the real and imaginary permittivity data are fitted simultaneously and $\varepsilon_{\infty}$, $\chi'_{U1}$, $n$, $\chi_{R0}$, $\tau$ and $\beta$ (or $\alpha$ and $\gamma$) are considered the adjustable parameters.

The KWW formula provides slightly but definitely better fitting than the HN one in spite of the fact that the number of adjustable parameters in the HN equation is larger and thus it is more versatile. This confirms that in PMN we really deal with the KWW relaxation.

The best-fit curves are shown in Fig. 2. The fit is quite good for high $T$, but below about 240 K the good fit can be obtained only in the low-$f$ part of the spectrum; at highest frequencies the experimental imaginary and real points deviate from the calculated curves upward and downward, respectively. This deviation can be explained by the existence of another (high-$f$) relaxation process which appears in the measurement frequency range only at low $T$. Indeed, this process has been revealed with the help of measurements at $f>10^{5}$ Hz [11]. To model the high-$f$ contribution we repeat fitting with an additional (fourth) term in Eq. 2 represented by the HN formula. In this case the quality of fit at low $T$ is significantly improved, confirming the existence of the high-$f$ relaxation contribution. However, due to the large number of adjustable parameters the results of fitting become ambiguous. On the other hand, in all possible variants of fitting the static limit of the additional term is found to be small ($\leq$ 2000) compared to the total measured $\varepsilon'$ and the most significant improvement of fit is achieved at the highest frequencies. Consequently the parameters $n$, $\beta$ and $\tau$ related to the low-$f$ relaxation processes remain practically the same as in the case of the initial fitting, provided $T$ is not too low.

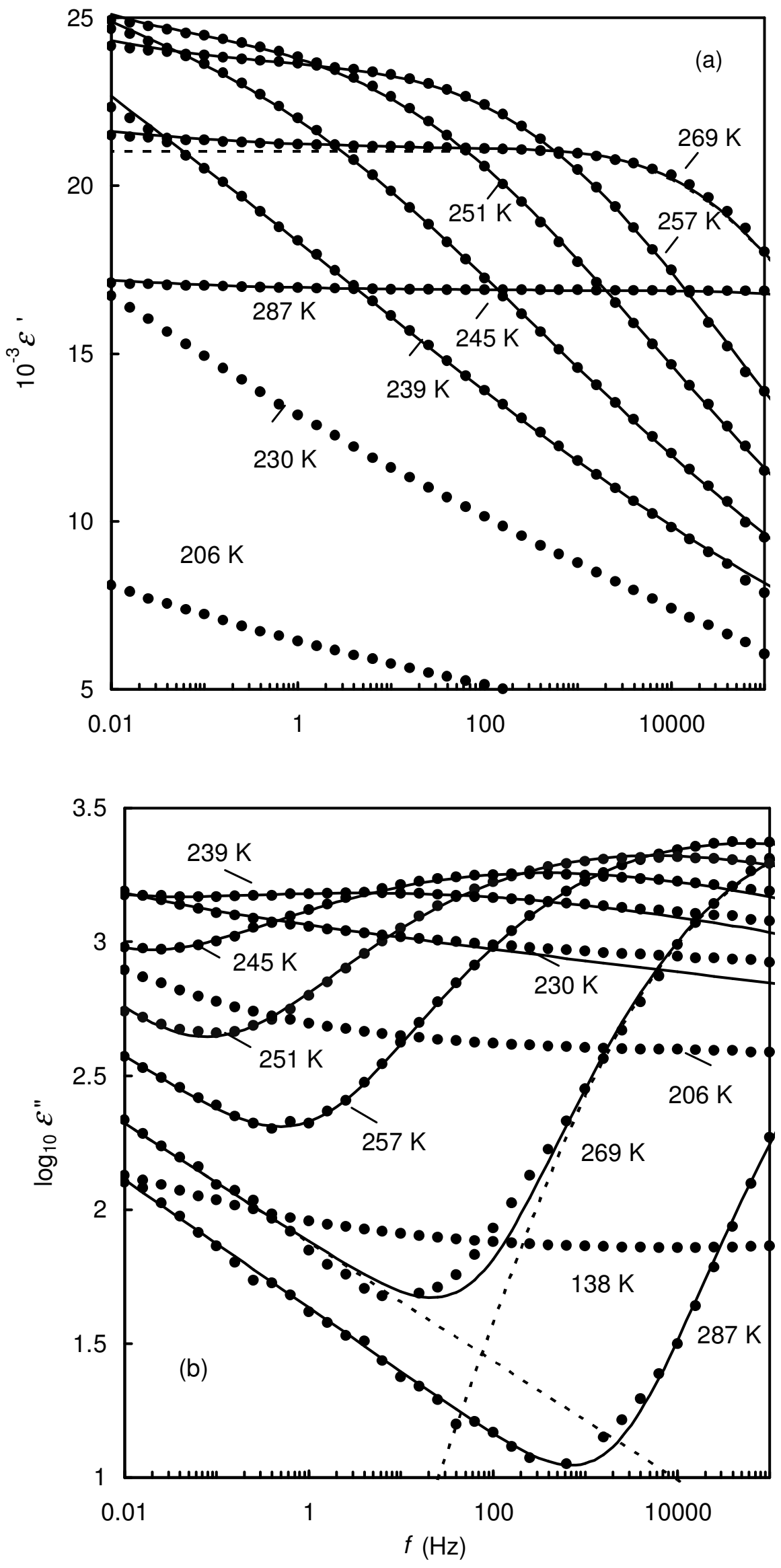

FIG. 2. Frequency dependences of (a) real and (b) imaginary part of $\varepsilon$ in PMN crystal at selected temperatures. Experimental data are shown by dots, solid lines represent fitting to Eqs (2)-(4). UR and CR components used for fitting at 269 K are shown by dashed lines.

We have found that in the whole temperature range where adequate fitting is possible $\tau$ follows the VF law (1) and the extrapolated value of $\tau$ diverges (relaxation frequency $f_{KWW}=1/(2\pi\tau)$ tends to zero) at $T_\tau$. This is shown in Fig. 4 plotted on such a scale that yields for the VF relation the straight line intersecting the abscissa at $T_\tau$. Furthermore, the other relaxation parameters derived from the fitting procedure also obey (see Fig. 4) the VF-type relations:

$$\beta = \beta_0 \exp\left[-E_\beta / \left(T - T_\beta\right)\right]; \tag{5}$$

$$n = n_0 \exp\left[-E_n / \left(T - T_n\right)\right]. \tag{6}$$

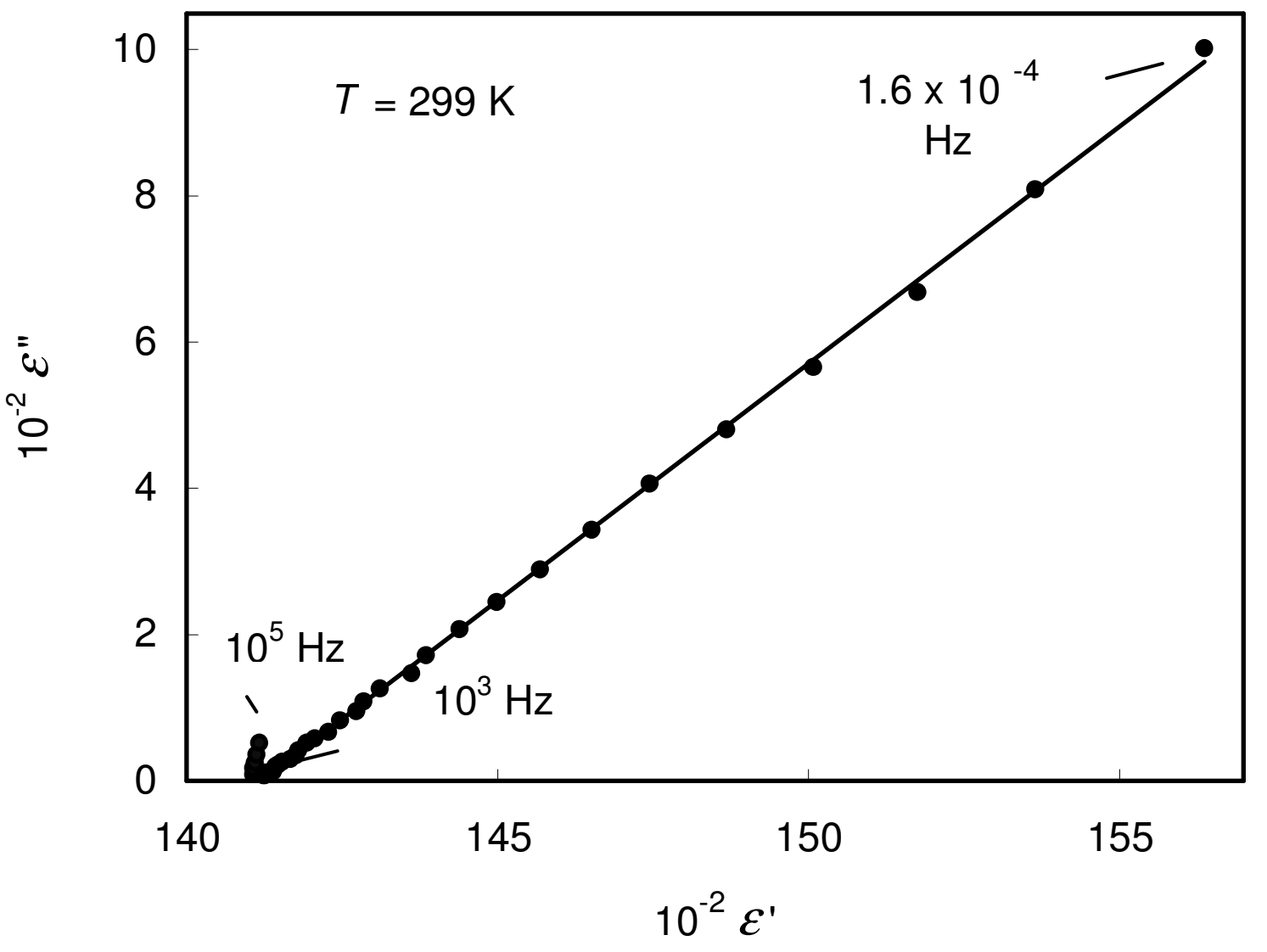

FIG. 3. Cole-Cole plot at 299 K. Experimental data (dots) were obtained at $f$=1.6×10$^{-4}$-10$^{5}$ Hz after cooling from 640 K with the rate of 1 K/min. Solid line is the fit of the low-frequency part of the spectrum to Eq. (3).

The best-fit parameters for Eqs. (1), (5), (6) are listed in Table 1 together with the parameters obtained for two other VF laws that describe the relations between the measuring $f$ and the position of the maximum on the temperature dependencies of the imaginary ($T_{mi}$) and real ($T_{mr}$) permittivity:

$$f = (2\pi\tau_{mi})^{-1}\exp\left[-E_{mi}/\left(T_{mi}-T_{fmi}\right)\right]; \qquad (7)$$

$$f = (2\pi\tau_{mr})^{-1}\exp\left[-E_{mr}/\left(T_{mr}-T_{fmr}\right)\right]. \qquad (8)$$

These two relations were reported previously for several relaxors, including PMN [6]. In the present work we confirm them in the range of 10$^{-2}$-10$^{5}$ Hz (for $T_{mr}$) and 10$^{2}$-10$^{5}$ Hz (for $T_{mi}$). At $f$<100 Hz the experimental $T_{mi}(f)$ data appear to deviate from Eq. (7) probably because of the increasing contribution of the UR relaxation process.

It is known that significant aging is characteristic of PMN at comparatively low $T$. As a result the measured $\varepsilon$ depends on the heating/cooling rate and on waiting time ($t_w$) the crystal has spent at a fixed $T$ before measurement. Our preliminary experiments with different $t_w$ show that aging influences $\tau$ but not $\beta$. Consequently, the characteristic temperature $T_\beta$, derived from the $\beta(T)$ dependences is not affected by aging phenomena.

Table 1. Parameters of the VF laws describing the behavior of relaxation parameters $\tau$, $\beta$, $n$, $T_{mi}$ and $T_{mr}$.

| Relaxation parameter | $\tau_0$, $\tau_{mi}$, $\tau_{mr}$ (s); $\beta_0$, $n_0$ | $E_\tau$, $E_\beta$, $E_n$, $E_{mi}$, $E_{mr}$ (K) | $T_\tau$, $T_\beta$, $T_n$, $T_{fmi}$, $T_{fmr}$ (K) |
|---|---|---|---|
| $\tau$ ($T$), Eq.(1) | 1.6×10$^{-14}$ | 810±60 | 213±2 |
| $\beta$ ($T$), Eq.(5) | 0.43±0.02 | 42±5 | 210±2 |
| $n$ ($T$), Eq.(6) | 0.87±0.02 | 7±3 | 211±15 |
| $T_{mi}$ ($f$), Eq.(7) | 6×10$^{-14}$ | 860±130 | 211±3 |
| $T_{mr}$ ($f$), Eq.(8) | 5×10$^{-14}$ | 880±30 | 221.6±0.6 |

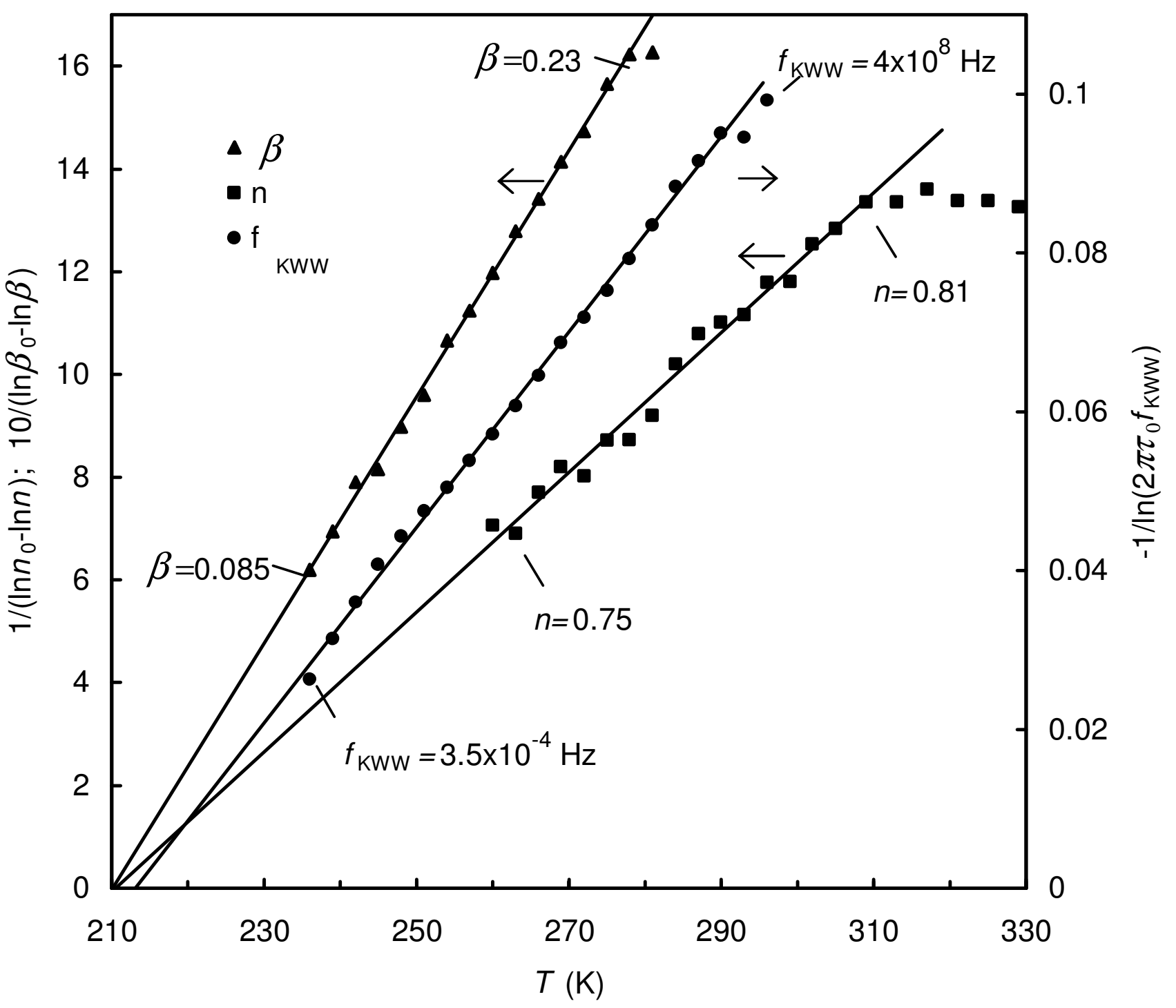

FIG. 4. Temperature dependences of the CS exponent $n$ and KWW relaxation parameters $\beta$ and $f_{KWW}=(2\pi\tau)^{-1}$. Solid lines are the fits to Eqs. (1), (5) and (6).

The temperature dependences of the UR susceptibility $\chi_{U1}\equiv\chi'_U(1\text{ Hz})$ and the static limit of the CR process ($\varepsilon_{R0}=\chi_{R0}+\varepsilon_\infty$) are presented in Fig. 1. In wide temperature intervals both dependences are found to obey a Lorenz-type function:

$$\varepsilon_{RA}/\varepsilon_{R0}=1+(T-T_{RA})^2/(2\delta_R^2), \quad (9)$$

$$\chi_{UA}/\chi_{U1}=1+(T-T_{UA})^2/(2\delta_U^2), \quad (10)$$

where the magnitude, temperature and width of the Lorenz peak are determined by the parameters $\varepsilon_{RA}=241\times10^2$ (or $\chi_{UA}=1.6\times10^2$ ), $T_{RA}$=245K (or $T_{UA}$=247 K) and $\delta_R$=46 K (or $\delta_U$=25 K), respectively.

Keeping in mind that many experimental data and theoretical models point to the existence of glass nonergodic state in PMN at low $T$ (see e.g. Ref [7] and Ref. [15] for a review) our results can be interpreted in the following way. Upon cooling from the ergodic phase τ diverges at $T_\tau$ which can be considered as glassy freezing temperature. At the same temperature the frequency interval of the KWW relaxation becomes infinitely wide ($\beta$ tends to zero). The freezing also leads to the VF behavior of $T_{mi}$ and $T_{mr}$, resulting in the same values for the first two parameters in Eqs. (1), (7), (8) (see Table 1); $T_{mi}$ is also the same as $T_\tau$ but $T_{mr}$ is slightly higher. The last conclusion resolves the long-standing dispute (see Refs. [6,16] for details) about the origin of relations (7), (8) in PMN.

As explained in Ref [12], the VF behavior for $n$ also implies freezing at $T_n$. The coincidence of $T_n$ with $T_\tau$ means that the subsystems of dipoles responsible for CR and UR(=CS) relaxations freeze out (or tend to be frozen) at the same temperature. At practically the same temperature other peculiarities were reported in PMN, including the sharp peak of hypersonic dumping (212 K) [17] and the Curie temperature (213 K) for the ferroelectric phase induced by an external electric field [15].

The temperatures $T_{RA}$ and $T_{UA}$ appear to be significantly higher than $T_\tau$. This implies that the maxima of $\varepsilon_{R0}(T)$ and $\chi_{U1}(T)$ are not due to the interactions among the species (dipoles) giving rise to the susceptibility (as in classical spin or dipole glasses); they are rather caused by the peaks in the temperature dependences of the number of the dipoles. The fact that $T_{RA} \approx T_{UA}$ suggests that the nature of the dipoles responsible for the CR and UR contributions is closely interrelated. It was proposed [18] that both contributions are associated with the characteristic polar nanoregions the number of which is known to vary considerably with temperature. The deviation of $\chi_{U1}(T)$ from the trend of Eq. (10) at low temperatures (see Fig. 1) is possibly due to the increasing dipolar interactions upon cooling.

It is instructive to compare PMN with PMNT in which the transition from the high-$T$ ergodic relaxor phase to the ferroelectric phase is known to take place at $T_C$ slightly below $T_{mr}$. One can expect that the differences between PMN and PMNT are general and reflect the distinction of canonical relaxors from the crystals with relaxor-to-ferroelectric phase transition. In PMNT the overlapping KWW and CS relaxations are also observed around $T_{mr}$ and Eqs. (6)-(9) hold [12,13]. However, $\beta$ does not change with $T$, $\tau$ increases dramatically on cooling but does not follow the VF or Arrhenius law [13], and $\chi_{U1}$ cannot be described by the continuous function (10), instead it diverges at $T_0 < T_{mr}$ as $\chi_{U1} \sim (T-T_0)^{-2}$ [13,18]. All this suggests that the dipolar subsystem responsible for the dielectric properties in PMNT is not subject to glassy freezing. The temperature $T_{fmi}$ derived from Eq. (7) can hardly be considered as freezing temperature; it rather arises from the non-critical evolution of relaxation spectrum as previously suggested [12] on the basis of the Tagantsev's theoretical prediction [16] implying the relation $T_{fmr} = T_{RA}$ really found in PMNT.

We believe that any successful theory of the dielectric response in relaxors should be able to derive the empirical formulae of this work. So far, no such theoretical justification is available.

This work was supported by U.S. ONR Grant No. N00014-99-1-0738.